\documentclass[aps,prl,showpacs,amsmath,twocolumn,amssymb,superscriptaddress,letterpaper]{revtex4}
\usepackage{graphicx,color}
\usepackage{amssymb}   % for math
\usepackage{amsmath}
\usepackage{mathdots}
\usepackage{epstopdf}
\usepackage{natbib}
\usepackage{hyperref}
\usepackage{epstopdf}
\usepackage{bm}

\begin{document}
%\title{Broken Time-reversal Symmetry Josephson in Tilted Weyl Semimetal }
\title{Quasi-one-dimensional Quantum Anomalous Hall Systems as New Platforms for Scalable Topological Quantum Computation}
\author{ Chui-Zhen Chen}
\affiliation{Department of Physics, Hong Kong University of Science and Technology, Clear Water Bay, Hong Kong, China. }
\author{ Ying-Ming Xie}
\affiliation{Department of Physics, Hong Kong University of Science and Technology, Clear Water Bay, Hong Kong, China. }
\author{ Jie Liu}
\affiliation{Department of Applied Physics, School of Science, Xian Jiaotong University, Xian 710049, China. }
\author{Patrick A Lee} \thanks{palee@mit.edu}
\affiliation{Department of Physics, Massachusetts Institute of Technology, Cambridge, MA, USA}
\author{K. T. Law} \thanks{phlaw@ust.hk}
\affiliation{Department of Physics, Hong Kong University of Science and Technology, Clear Water Bay, Hong Kong, China. }

\begin{abstract}
Quantum anomalous Hall insulator/superconductor heterostructures emerged as a competitive platform to realize topological superconductors with chiral Majorana edge states as shown in recent experiments [He et al. Science {\bf 357}, 294 (2017)]. However, chiral Majorana modes, being extended, cannot be used for topological quantum computation. In this work, we show that quasi-one-dimensional quantum anomalous Hall structures exhibit a large topological regime (much larger than the two-dimensional case) which supports localized Majorana zero energy modes. The non-Abelian properties of a cross-shaped quantum anomalous Hall junction is shown explicitly by time-dependent calculations. We believe that networks of such quasi-one-dimensional quantum anomalous Hall systems can be easily fabricated for scalable topological quantum computation.
\end{abstract}
%\pacs{}

\maketitle

{\emph {Introduction.}}  The search for Majorana zero energy modes which obey non-Abelian statistics has been one of the most exciting areas of research in the past decade \cite{Kitaev2003,Nayak2008}.
It was first proposed that two-dimensional $p+ip$-wave superconductors support chiral Majorana edge states \cite{Read,Gurarie,Stone}, while one-dimensional $p$-wave superconductors support localized Majorana zero energy modes at the ends of the superconductors \cite{Kitaev2001}. Several experimentally promising schemes to engineer effective $p+ip$ topological superconductors which support Majorana modes have been proposed \cite{Fu2008,Sun2016}. First, Fu and Kane showed that the vortex cores of superconducting surface states of topological insulators host Majorana fermions \cite{Fu2008}. Spin dependent zero bias conductance peaks possibly associated with Majorana fermion in superconducting topological superconductors have been observed \cite{JamesHe,Sun2016}. Second, it was also suggested that Majorana modes can exist as end states of semiconductor wires in proximity to superconductors \cite{Lutchyn,Alicea1,Oreg}. Zero bias conductance peaks have been reported as signatures of Majorana fermions in these semiconductor-wire/superconductor heterostructures \cite{Mourik,Ronen,Deng}. More recently, the observation of the predicted $2e^2/h$ conductance peak \cite{Law,Wimmer}  provides strong evidence that Majorana fermions have been observed in this system \cite{HaoZhang}. However, it remains challenging to fabricate scalable Majorana networks based on the semiconductor schemes \cite{Karzig}.

\begin{figure}[bth]
\centering
\includegraphics[scale=0.5, bb =0 10 450 350, clip=true]{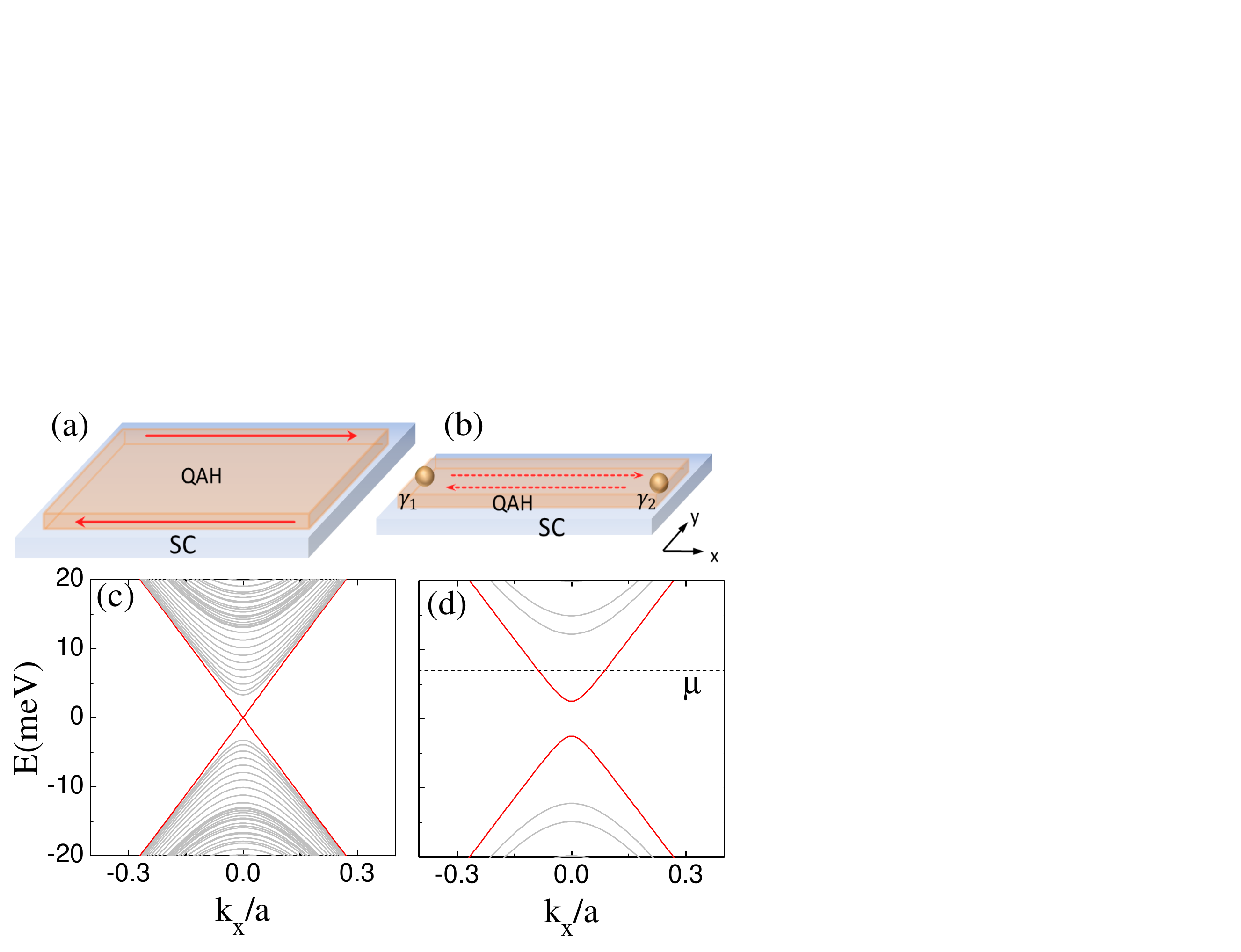}
\caption{(Color online).  (a) Schematic plot of a QAH insulator with two chiral edge modes (red arrows) in proximity to an s-wave superconductor (SC).
(b) The QAH insulator is narrowed down such that the two chiral edge modes are coupled to form a single helical conducting channel (red dash arrows), inducing superconductivity gives rise to two Majorana modes $\gamma_1$ and $\gamma_2$.
(c) The energy spectrum of quasi-1D QAH insulator with width  $L_y =800$nm. The two edge modes on opposite sides are approximately decoupled as the $L_y$ is much larger than the localization length of the chiral edge modes.
(d) The energy spectrum with $L_y =100$nm. The dashed line indicates the possible location of the chemical potential in the topological regime. A gap will open at the chemical potential upon proximity coupling to a superconductor.
A tight-binding version of $H_{QAH}$ is used for (c) and (d) and the parameters are: $\hbar v_F=3$eV$\cdot$\AA, $m_0=-5$meV, $m_1=15$eV$\cdot$\AA$^2$, $M_z=8$meV.
\label{fig1} }
\end{figure}

A recent paper reported the realization of the effective $p+ip$ wave superconductor with chiral Majorana edge states in quantum anomalous Hall/superconductor  (QAH/SC) heterostructures, base on the observation of half-quantized conductance plateaus in two-terminal transport experiments \cite{He} as predicted previously \cite{Wangjing,Qi,Chung}. Unfortunately, chiral Majorana modes, being extended, cannot be used for topological quantum computation. Here, we suggest using quasi-one-dimensional (quasi-1D) QAH/SC heterostructures to realize localized Majorana zero energy modes. Importantly, the topological regime of this geometry is much larger than the two-dimensional chiral topological superconducting case, which requires the superconducting pairing gap to be larger than the bulk gap of the QAH system. Moreover, the experimental parameters of Cr or V doped (Bi,Sb)$_2$Te$_3$ QAH thin films are highly tunable by external magnetic fields and gating \cite{Yu,Chang,Checkelsky,Kou,Bestwick,Feng,Chang2,Kandala}, and complicated device geometry can be achieved by standard fabrication techniques. Therefore, the system provides a promising platform for braiding Majorana zero energy modes and scalable quantum computations.

In the following sections, we first demonstrate that quasi-1D QAH systems in proximity to an s-wave superconductor exhibits a large, experimentally accessible, topological regime, which supports localized Majorana zero energy modes. When the quasi-1D QAH system is narrow compared with the localization length of the edge modes, the two chiral edge modes of QAH are coupled to form a single helical conducting channel as shown in Fig.1. When superconductivity is induced on this single conducting channel, an effective $p$-wave superconductor is realized similar to the Rashba semiconductor wire systems proposed previously \cite{Lutchyn,Alicea1,Oreg}. Importantly, the edge modes in quasi-1D QAH are well separated from the bulk states, giving rise to a topological regime of about 10meV which is 1 order of magnitude larger than the topological regimes in semiconductor wires \cite{Mourik,Liujie}. Finally, we performed a time-dependent calculation to show the non-Abelian braiding dynamics of Majorana modes in a cross-shaped QAH junction. Such QAH structures can readily be extended to a scalable network and provide new platforms for topological quantum computation.

{\emph {Model Hamiltonian.}} The effective Hamiltonian of QAH insulator can be written as \cite{Yu,Wangjing}
\begin{eqnarray}
% \nonumber to remove numbering (before each equation)
  H_{QAH}&=&\sum_{\bf k} \Psi_{\bf k}^{\dagger}[\hbar v_F k_y \sigma_x\tau_z - \hbar v_F k_x \sigma_y\tau_z \!+\! m({\bf k}) \tau_x \nonumber\\
  && + M_z \sigma_z] \Psi_{\bf k}.
\end{eqnarray}

Here, $\Psi_{\bf k}=[\psi_{{\bf k}t \uparrow}, \psi_{{\bf k}t \downarrow},\psi_{{\bf k}b \uparrow},\psi_{{\bf k}b\downarrow}]$
is a four-component electron operator with the momentum ${\bf k}$, where $t$ ($b$) denotes the top (bottom) layer of topological insulator surface and $\uparrow$ ($\downarrow$) denotes the spin index.
The Pauli matrices $\sigma_{x,y,z}$  operate on spin space and $\tau_{x,z}$ operate on the layer index. Fermi velocity of the surface states is denoted by $v_F$ and $m({\bf k}) = m_0 - m_1(k_x^2 + k_y^2)$ describes the effective coupling between the top and bottom layers. $M_z$ denotes spin splitting in z-direction due to the magnetic doping in topological insulator and external magnetic field.
To be specific, we set $m_0=-5$meV and $m_1=15$eV$\cdot$\AA$^2$ \cite{Xue}.
When $M_z<-|m_0|$ and $M_z>|m_0|$ , the system is in the QAH phase with Chern number $N=1$ and $-1$ respectively.
On the other hand, the system is a trivial insulator for $-|m_0|<M_z<|m_0|$. We have shown recently that this effective Hamiltonian can accurately describe the experimental data in the conductance measurements of QAH-superconductor heterostructures \cite{CZChen}.
%We note that the $T H(M_z) T^{-1}=H(-M_z)$ with time-reversal operator $T=i\sigma_y \mathcal{K}$ and the complex conjugation $\mathcal{K}$.

%To study quasi-1D QAH Hamiltonian, we rewrite $H$ in real space in y direction
%as
%\begin{eqnarray}
% \nonumber to remove numbering (before each equation)
%  \mathcal{H}&=&\!\!\sum_{k_x} \!\! \sum_{n_y=1}^{N_y} \Psi_{k_x,n_y}^{\dagger}[  -\hbar v_F k_x \sigma_y \tau_z +m_x \tau_x + M_z \sigma_z] \Psi_{k_x,n_y} \nonumber\\
%  && + {\Psi}_{k_x,n_y}^{\dagger} (\frac{\hbar v_F}{2ia} \sigma_x \tau_z +\frac{m_1}{2a^2}\tau_x )\Psi_{k_x,n_y+1} + h.c.
%\end{eqnarray}

As described earlier, when the sample is narrowed down to a quasi-1D limit ( when the width of the sample is comparable to the localization length of the chiral edge modes), as shown in Fig.1b, the two chiral edge states can couple to form a helical channel.
In the calculations for Fig.\ref{fig1}c and d, the localization length $\xi \approx \hbar v_f/E_{g}$ of two edge modes is about $100$nm with a bulk gap of $E_{g}=3$meV.
When the width of QAH system is much larger than the localization length, for example, $L_y=800$nm$\gg\xi$, the two edge modes on the opposite sides are not coupled [see Fig.\ref{fig1}(c)].
However, when we reduce the width of the QAH system to be comparable to the localization length of $100$nm, the two edge modes are coupled and open up a sizable gap of about 5meV. The resulting energy spectrum is shown in Fig.\ref{fig1}(d). Due to the increased confinement, the bulk states are pushed to higher energy which leaves a well separated helical channel from the rest of the states. This important feature gives rise to a large topological superconducting regime ($\approx$ 10meV) in the quasi-1D QAH/SC heterostructure as discussed below.
%The overlap of two edge states makes it possible to have an obvious proximity pairing gap when the chemical potential cut the single chiral edge modes if we put a superconductor on this qusi one d %QAH.
 %Importantly, the edge modes in QAH are well separated from the bulk states, giving rise to large topological superconductor region
%when it is in proximate to a superconductor in the following.

{\emph {Localized Majorana modes in QAH/SC heterostructures.}} In proximity to an s-wave superconductor, the quasi-1D
QAH/SC heterostructure Hamiltonian can be expressed as \cite{Wangjing}
\begin{eqnarray}
% \nonumber to remove numbering (before each equation)
H_{BdG} \!\! &=&\!\!\! \sum_{k_x}\!\sum_{n_y=1}^{N_y} \!\Phi_{k_x,n_y}^{\dagger}[(m_x \tau_x \! -\! \hbar v_F k_x \sigma_y \tau_z  \!+\! M_z \sigma_z\!-\!\mu ) s_z \nonumber\\
  && +\Delta\sigma_y\frac{\tau_{z} + I}{2} s_y] \Phi_{k_x,n_y}+ \Phi_{k_x,n_y}^{\dagger}(\frac{1}{2ia}\hbar v_F \sigma_x \tau_z    \nonumber\\
  &&  +\frac{1}{a^2}m_1\tau_x s_z )\Phi_{k_x,n_y+1} + h.c.
\end{eqnarray}
where $\Phi_{k_x,n_y}$=$[\Psi_{k_x,n_y}$,$\Psi_{-k_x,n_y}^{\dagger}]^T$,
and $\Psi_{k_x,n_y}^\dagger$ creates an electron at $n_y$ site with momentum $k_x$.
$\Delta$ is the paring potential on the top surface due to the superconductor and $I$ is the unit matrix.
$\mu$ is chemical potential. $m_x = m_0 - m_1(k_x^2+2/a^2)$ and the size $L_y = N_y a$ with lattice constant $a$.
$s_{x,y,z}$ are Pauli matrices defined on particle-hole space.
The Hamiltonian respects a particle-hole symmetry $\mathcal{P} H_{BdG}(k_x) \mathcal{P}^{-1} = -H_{BdG}(- k_x)$ and a time-reversal like symmetry
$\mathcal{T} H_{BdG}(k_x) \mathcal{T}^{-1} = H_{BdG}(- k_x)$ with $\mathcal{P}=s_x\mathcal{K}$ and $\mathcal{T}=\mathcal{UK}$.
Here $\mathcal{K}$ is the complex conjugate operator
and the $N_y\times N_y$ anti-diagonal matrix $\mathcal{U}_{ij}=\delta_{i+j, N_y+1}$ is defined in real space with Kronecker delta $\delta_{i,j}$.
As a consequence, $H_{BdG}$ satisfies a chiral symmetry $\mathcal{C} H_{BdG}(k_x) \mathcal{C}^{-1} = H_{BdG}(- k_x)$,
and  the system is in BDI class with $\mathcal{T}^2=1$ and $\mathcal{P}^2=1$, where $\mathcal{C}=\mathcal{PT}=\mathcal{U}s_x$. Therefore, the topological properties of the system is characterized by a topological invariant $N_{BDI}$ for BDI class Hamiltonians \cite{Schnyder,Tewari}.

\begin{figure}[tbh]
\centering
\includegraphics[scale=0.35, bb =0 0 1000 350, clip=true]{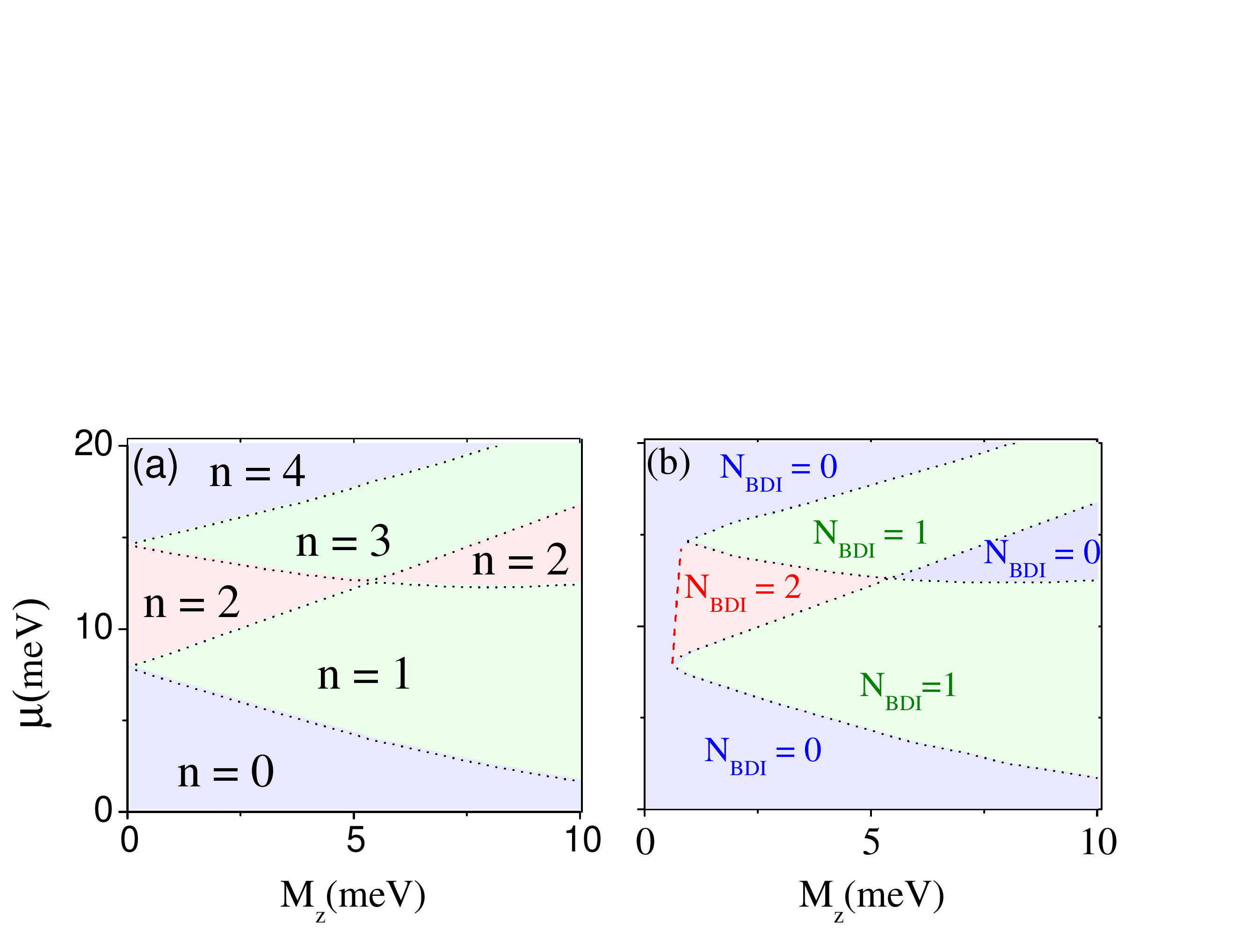}
\caption{(a)  The number $n$ denotes the number of bands of the quasi-1D QAH (without superconductivity) cut by the Fermi energy as a function of chemical potential $\mu$ and effective Zeeman field $M_z$. The width and parameters of the quasi-1D system are the same as in Fig.1d. (b) The topological invariant $N_{BDI}$ in the superconducting phase as a function of  $\mu$ and $M_z$. The parameters are the same as (a) except that a pairing potential $\Delta=1$meV is introduced on the top surface.
\label{fig2} }
\end{figure}

In Fig.\ref{fig2}(a), we show the number of bands $n$ of the quasi-1D QAH (without superconductor) cut by the Fermi energy $\mu$ with Zeeman field $M_z$.
In the absence of Zeeman field $M_z=0$, the system has time-reversal symmetry $\mathfrak{T}=i\sigma_y$ giving rise to two folder Kramer's degeneracies.
The Fermi energy can only cut through an even number of bands $n=0,2,4$.
With increasing $M_z$, each one-dimensional degenerate band splits into two branches and there can be an odd number of bands at the Fermi energy. As shown in Fig.2(b), the system is topological when there is an odd number of bands at the Fermi energy. Indeed, there is a $N_{BDI}=2$ phase when two subbands are partially occupied. In this phase, there are two Majorana fermions at each end of the wire which are protected by the chiral symmetry. Unfortunately, disorder breaks this chiral symmetry and the two Majorana modes at one end of the wire will couple into a trivial fermionic mode. Therefore, only the $N_{BDI}=1$ phase is topologically non-trivial in the presence of disorder as the single Majorana can be protected by the particle-hole symmetry alone. It is important to note that the $N_{BDI}=1$ topological region is very wide, which can be tuned by chemical potential $\mu$ or Zeeman field $M_z$. For example, when $M_z>m_1=5$meV, which can be easily reached by tuning the external magnetic field, the lowest band originating from the QAH edge mode is well separated from the second lowest band originating from the QAH bulk band. This gives rise to a topological regime of about $10$meV wide in terms of chemical potential as shown in Fig.\ref{fig2}(b). The topological regime is about ten times larger than that of semiconductor wire systems \cite{Mourik,Liujie}. \emph{This regime is also much wider than the two-dimensional chiral topological regime which requires $\Delta$ to be larger than $M_z$} \cite{Qi,Wangjing}. Here, there are no such requirements. Therefore, the quasi-1D QAH structure hosts a large  topological regime which is readily accessible experimentally.
%Besides, the QAH edge modes can be created without magnetic external field \cite{Chang,Checkelsky,Kou,Bestwick,Feng,Chang2},  which avoiding to sabotage the superconductivity of the %heterostructures.

\begin{figure}[tbh]
\centering
\includegraphics[scale=0.37, bb =00 20 900 500, clip=true]{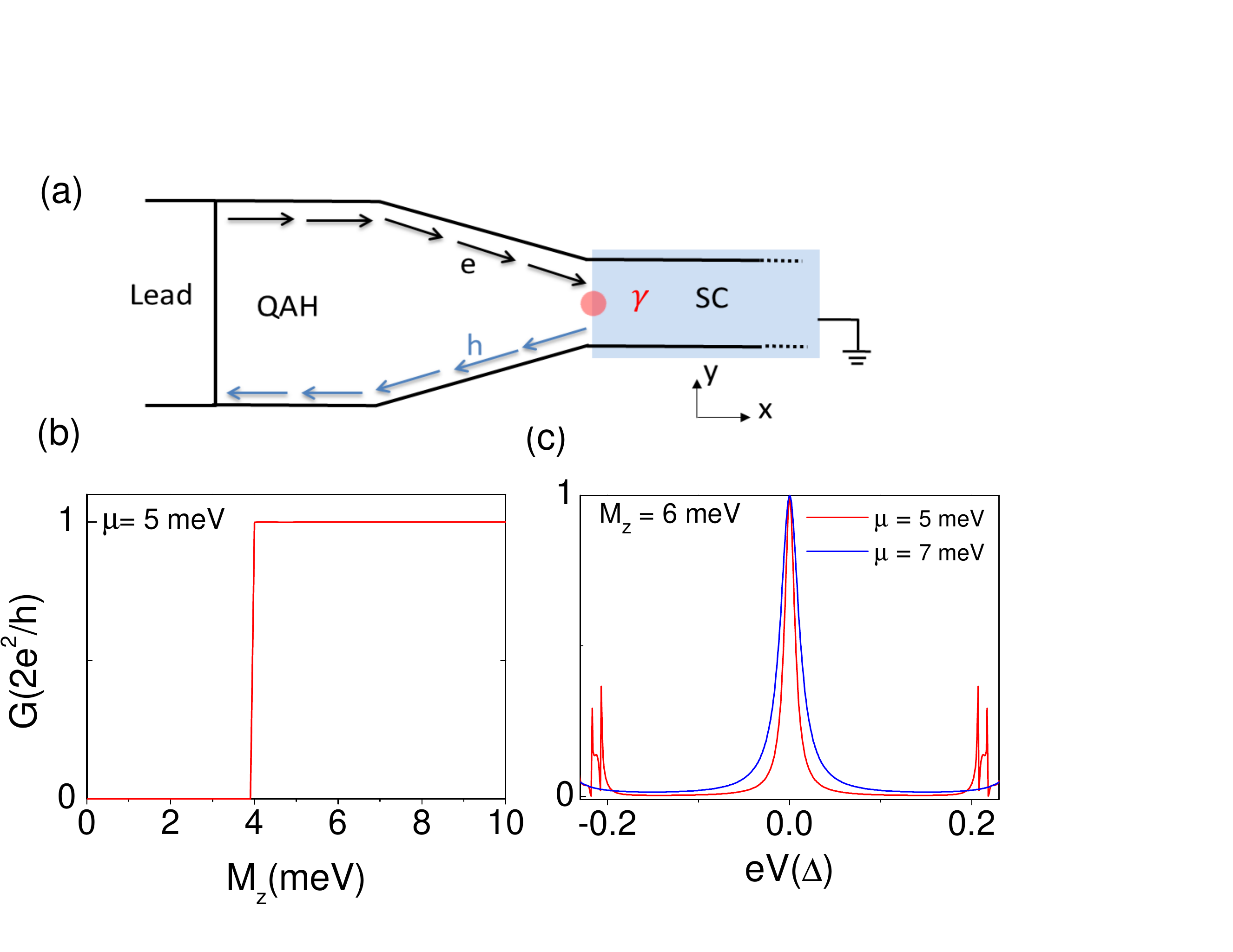}
\caption{(a) Schematic plot of two-terminal device for the detection of Majorana zero modes. A lead is attached to a QAH system while the narrow region of the QAH system is in proximity to an s-wave superconductor to create the Majorana zero energy mode. An electron mode from the edge state of the QAH system (black arrows) is reflected as a hole mode (blue arrows) by a Majorana mode $\gamma$ (red dot).
(b) The zero bias conductance $G$ from the lead to the superconductor as a function of Zeeman field $M_z$ with chemical potential $\mu=5$meV.
(c) The two terminal conductance versus electric voltage $eV$ for different $\mu$ with $M_z=6$meV and $\Delta=1$meV. Other parameters are the same as those in Fig.\ref{fig2}(b).
\label{fig3} }
\end{figure}

To detect the Majorana modes, one may attach a macroscopic lead to the QAH system as shown in Fig.3a. When the QAH system possesses chiral fermionic modes, an electron mode (black arrows) comes into the QAH and is reflected as a hole mode (blue arrows) by a Majorana mode localized at the end of the QAH/SC heterostructure. This gives rise to the resonant Andreev reflections predicted previously \cite{Law,Wimmer}. The conductance of the system is evaluated by recursive Green's function methods numerically \cite{Liujie,PALee}. As expected, by tuning the QAH system from a trivial insulating state to a QAH state, a conductance plateau is observed in Fig.3(b). Deep in the topological regime with fixed $M_z$, a zero bias conductance peak is observed as shown in Fig.3 (c).

\begin{figure}[tbh]
\centering
\includegraphics[scale=0.24, bb =0 0 1100 420, clip=true]{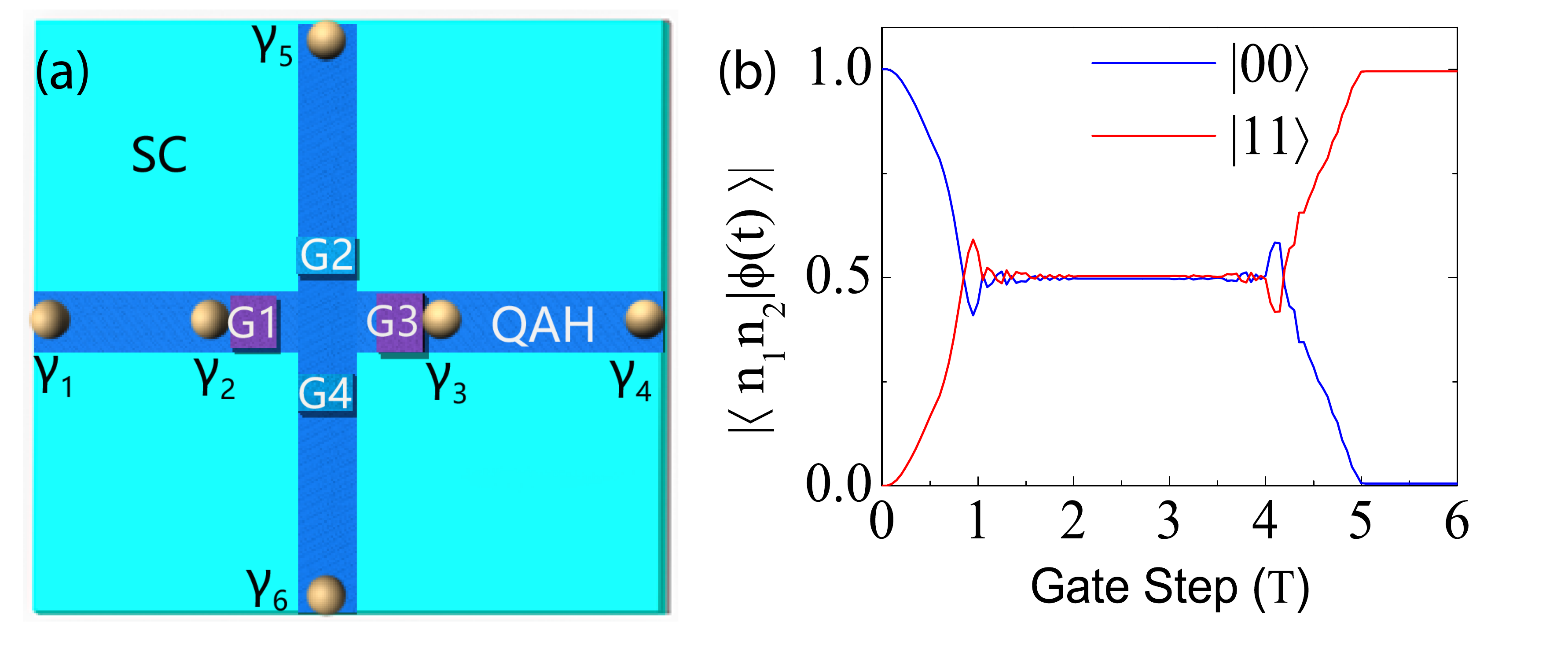}
\caption{ (a) Schematic plot of a cross-shaped QAH system in proximity to a superconductor with four tunable gates (G1-G4) on the top.
Initially, when the G1 and G3 are turned on with G2 and G4 are turned off, the system is divided into three topologically nontrivial parts with six Majorana modes ($\gamma_1$-$\gamma_6$). (b) The projection of wavefunction $\phi(t)$ onto the initial eigenstate $|n_1n_2\rangle$ as a function of time, where $\phi(t) = \mathbb{T} e^{i\int dt H_{C}/\hbar} |n_1n_2\rangle$ with the time-ordered operator $\mathbb{T}$ and $H_{C}$ is a realistic tight-binding model describing the cross-shaped QAH junction. The six-stage process of braiding $\gamma_2$ and $\gamma_3$ is described in the main text. The final states are orthogonal to the initial states after braiding and non-Abelian statistics is clearly demonstrated. The total braiding time for the six steps is $6T=14400\hbar/\Delta \approx 9.5$ns. The details of the calculations can be found in the Supplemental Material \cite{SM}.
\label{fig4} }
\end{figure}

{\emph {Non-Abelian braiding dynamics in QAH structure}}
On the basis of the large topological region in the QAH structure above, we suggest to use a cross-shaped junction of the QAH/SC for braiding Majorana modes \cite{Amorim}. This cross-shaped junction scheme does not require gating the bulk states and provides simpler braiding processes compared to the T-junction scheme \cite{Alicea}. Importantly, we expect that a scalable network for quantum computation can be easily fabricated by standard techniques from two dimensional QAH systems. In Fig.\ref{fig4}(a), the cross-shaped QAH is fabricated on the superconductor with four gates (G1-G4) on the top. Initially, the G1 and G3 are turned on (potential barriers are created) while G2 and G4 are turned off (no gating potential is applied), and the QAH/SC are divided into three topological nontrivial parts such that there are six Majorana modes in the system ($\gamma_1$-$\gamma_6$) as shown in Fig.\ref{fig4}(a).

The effective Hamiltonian of the QAH cross-shaped junction can be written as
\begin{eqnarray*}
% \nonumber to remove numbering (before each equation)
  H_{eff} &=& i\epsilon_1\gamma^{\vphantom{\dagger}}_1 \gamma^{\vphantom{\dagger}}_2 + i\epsilon_2\gamma^{\vphantom{\dagger}}_4 \gamma^{\vphantom{\dagger}}_3
\end{eqnarray*}
where $\epsilon_{1,2}$ are the coupling energies of Majorana modes. Here we ignore $\gamma_5$ and $\gamma_6$ in Hamiltonian, since they are well separated.
Now, the system contains two fermionic modes $c_1=\gamma_1 + i \gamma_2$ and $c_2=\gamma_4 + i \gamma_3$,
and  the low energy Hilbert space contains four qubit states $|n_1n_2\rangle$ with particle number $n_{1,2}=0,1$.  The ground state is defined as $c_{1}|00\rangle=c_{2}|00\rangle=0$.
 The braiding operator $B(\gamma_2,\gamma_3)=\exp(\frac{\pi}{4}\gamma_3\gamma_2)$, transforms the Majorana modes as
 $\gamma_2\rightarrow\gamma_3$ and  $\gamma_3\rightarrow-\gamma_2$ \cite{Ivanov}.
 Exchanging $\gamma_2$ and $\gamma_3$ twice would lead to a sign change $\gamma_2\rightarrow-\gamma_2$ and  $\gamma_3\rightarrow-\gamma_3$,
 and then $c_1=\gamma_1 + i \gamma_2\rightarrow c_1^\dagger$ and $c_2=\gamma_4 + i \gamma_3 \rightarrow c_2^\dagger$,
 As a result, if the initial state of the system is $|00\rangle$, the final state $|11\rangle\rightarrow B^2(\gamma_2,\gamma_3)|00\rangle$ becomes orthogonal to the initial state
after exchanging $\gamma_2$ and $\gamma_3$ twice.

In the time-dependent simulations using a realistic Hamiltonian, $\gamma_2$ and $\gamma_3$ can be exchanged by taking three steps: first, move $\gamma_2$ upward by turning off G1 and then turning on G2;
second, move $\gamma_3$ to the left by turning off G3 and then turning on G1; finally, move $\gamma_2$ to the right by turning off G2 and then turning on G3.
Exchanging $\gamma_2$ and $\gamma_3$ twice yields a full braiding process.
In Fig.\ref{fig4}(b), when the initial state is set to be $|00\rangle$ (blue line),
we reach the final state $|11\rangle$ (red line) after braiding $\gamma_2$ and $\gamma_3$ in the six-step gating process as described above. Indeed, the final state $|11\rangle$ is orthogonal to the initial state $|00\rangle$ after the braiding process. This clearly justifies the non-Abelian nature of the Majorana braiding process and thus QAH/SC heterostructures can provide new platforms for topological quantum computation. In the above simulation, time-evolution operator $U = \mathbb{T} e^{i\int dt H_{C}/\hbar}$ is used where $\mathbb{T}$ is time ordered operator and $H_{C}$ is a realistic real space tight-binding Hamiltonian describing the cross-shaped QAH junction. The final time is set to be $6T=14400\hbar/\Delta \approx 9.5$ns with $\Delta=1$meV. The details of the calculations can be found in the Supplemental Material \cite{SM}.

{\emph{Conclusion and discussion.} In conclusion, we have shown that quasi-1D QAH/SC heterostructures exhibits a large, experimentally accessible, topological regime which supports localized Majorana zero energy modes. This quasi-1D topological superconductor phase does not require the superconducting paring gap $\Delta$ to be larger than the QAH gap,
as in the case of chiral superconducting phase in the QAH/SC heterostructures \cite{Qi, Wangjing,He}. Therefore, the results discussed in this work does not depend on the origin of the half-quantized plateaus observed recently by He et al.  \cite{He} which is under intense debate \cite{CZChen,WJi,YHuang,BLian}. Moreover, due to hysteresis of QAH systems, the edge modes in QAH systems and the Majorana modes can be created without applying external magnetic fields \cite{Chang,Checkelsky,Kou,Bestwick,Feng,Chang2}. Finally, we explicitly demonstrated the non-Abelian braiding dynamics of Majoranas in a quasi-1D QAH structure using realistic parameters. We believe that cross-shaped junction and even more complicated structures can be made by experimentalists for quantum computation in the near future.

{\emph{Acknowledgement.}} We thank Junwei Liu and Noah Yuan for illuminating discussions. KTL acknowledges the support of HKRGC, Croucher Foundation and Dr. Tai-chin Lo Foundation through C6026-16W, 16303014, 16324216, 16307117 and Croucher Innovation Grant. JL is supported by NSF-China under Grant Nos.11574245. PAL acknowledges the support by DOE grant FG02-03 ER46076 and thanks the hospitality of the Institute for Advanced Studies at HKUST.


\begin{thebibliography}{QAH_SC}
\bibitem{Kitaev2003} A. Kitaev, Ann. Phys. (N.Y.) {\bf303}, 2 (2003).
\bibitem{Nayak2008} C. Nayak, S. H. Simon, A. Stern, M. Freedman, and S. Das Sarma, Rev. Mod. Phys. {\bf80}, 1083 (2008).
\bibitem{Read} N. Read and D. Green, Phys. Rev. B {\bf61}, 10267 (2000).
\bibitem{Gurarie} V. Gurarie, L. Radzihovsky, and A. V. Andreev, Phys. Rev. Lett. {\bf94}, 230403 (2005).
\bibitem{Stone} M. Stone and S. B. Chung, Phys. Rev. B {\bf73}, 014505 (2006).
\bibitem{Kitaev2001}A. Y. Kitaev, Phys. Usp. {\bf44}, 131 (2001).
\bibitem{Fu2008} L. Fu and C. L. Kane, Phys. Rev. Lett. {\bf100}, 096407 (2008).
\bibitem{Sun2016} H. -H. Sun et al., Phys. Rev. Lett. {\bf116}, 257003 (2016).
\bibitem{JamesHe}J. J. He, T. K. Ng, P. A. Lee, and K. T. Law, Phys. Rev. Lett. {\bf112}, 037001 (2014).
\bibitem{Lutchyn} R. M. Lutchyn, J. D. Sau, and S. Das Sarma, Phys. Rev. Lett. {\bf105}, 077001 (2010).
\bibitem{Alicea1} J. Alicea, Phys. Rev. B {\bf 81}, 125318 (2010).
\bibitem{Oreg}Y. Oreg, G. Refael, and F. von Oppen, Phys. Rev. Lett. {\bf 105},177002 (2010).
\bibitem{Mourik} V. Mourik, K. Zuo, S. M. Frolov, S. R. Plissard, E. P. A. M. Bakkers, and L. P. Kouwenhoven, Science {\bf336}, 1003 (2012).
\bibitem{Ronen}A. Das, Y. Ronen, Y. Most, Y. Oreg, M. Heiblum, and H. Shtrikman, Nat. Phys. {\bf8}, 887 (2012).
\bibitem{Deng}M. T. Deng, C. L. Yu, G. Y. Huang, M. Larsson, P. Caroff, and H. Q. Xu, Nano Lett. {\bf 12}, 6414 (2012).
\bibitem{Law} K. T. Law, P. A. Lee, and T. K. Ng, Phys. Rev. Lett. {\bf103}, 237001 (2009).
\bibitem{Wimmer}M. Wimmer, A. R. Akhmerov, J. P. Dahlhaus, and C. W. J. Beenakker, New J. Phys. 13, 053016 (2011).
\bibitem{HaoZhang} We thank Leo Kouwenhoven and Hao Zhang for the private communication.
\bibitem{Karzig}T. Karzig, C. Knapp, R. Lutchyn, P. Bonderson, M. Hastings, C. Nayak, J. Alicea, K. Flensberg, S. Plugge, Y. Oreg, C. Marcus, and M. H. Freedman, Phys. Rev. B {\bf95}, 235305 (2017).
\bibitem{He}Q. L. He, L. Pan, A. L. Stern, E. Burks, X. Che, G. Yin, J. Wang, B. Lian, Q. Zhou, E. S. Choi, K. Murata, X. Kou, T. Nie,
Q. Shao, Y. Fan, S.-C. Zhang, K. Liu, J. Xia, and K. L. Wang, Science {\bf 357}, 294 (2017).
\bibitem{Qi}X. L. Qi, T. L. Hughes, and S. C. Zhang, Phys. Rev. B {\bf82}, 184516 (2010).
\bibitem{Chung} S. B. Chung, X. L. Qi, J. Maciejko, and S. C. Zhang, Phys. Rev.B {\bf83}, 100512(R) (2011).
\bibitem{Wangjing}  J. Wang, Q. Zhou, B. Lian, and S. C. Zhang, Phys. Rev. B {\bf92}, 064520 (2015).
\bibitem{Yu} R. Yu, W. Zhang, H. J. Zhang, S. C. Zhang, X. Dai, and Z. Fang, Science {\bf329}, 61 (2010).
\bibitem{Chang} C.-Z. Chang, J. Zhang, X. Feng, J. Shen, Z. Zhang, M. Guo, K. Li, Y. Ou, P. Wei, L.-L. Wang, Z.-Q. Ji, Y. Feng, S. Ji, X. Chen,
                J. Jia, X. Dai, Z. Fang, S.-C. Zhang, K. He, Y. Wang, L. Lu, X.-C. Ma, and Q.-K. Xue, Science {\bf 340}, 167 (2013).
\bibitem{Checkelsky} J. G. Checkelsky, R. Yoshimi, A. Tsukazaki, K. S. Takahashi, Y. Kozuka, J. Falson, M. Kawasaki, and Y. Tokura, Nat. Phys. {\bf 10}, 731 (2014).
\bibitem{Kou} X. Kou, S.-T. Guo, Y. Fan, L. Pan, M. Lang, Y. Jiang, Q. Shao, T. Nie, K. Murata, J. Tang, Y. Wang, L. He, T.-K. Lee, W.-L. Lee, and K. L. Wang, Phys. Rev. Lett. {\bf 113}, 137201 (2014).
\bibitem{Bestwick} A. J. Bestwick, E. J. Fox, X. Kou, L. Pan, K. L. Wang, and D. Goldhaber-Gordon, Phys. Rev. Lett. {\bf 114}, 187201 (2015).
\bibitem{Feng}Y. Feng, X. Feng, Y. Ou, J. Wang, C. Liu, L. Zhang, D. Zhao, G. Jiang, S. C. Zhang, K. He, X. Ma, Q. K. Xue, and Y. Wang, Phys. Rev. Lett. {\bf 115}, 126801 (2015).
\bibitem{Chang2} C.-Z. Chang, W. Zhao, D. Y. Kim, H. Zhang, B. A. Assaf, D. Heiman, S.-C. Zhang, C. Liu, M. H. W. Chan, and J. S. Moodera, Nat. Mater. {\bf 14}, 473 (2015).
\bibitem{Kandala} A. Kandala, A. Richardella, S. Kempinger, C.-X. Liu, and N. Samarth, Nat. Commun. {\bf6}, 7434 (2015).
\bibitem{Liujie} J. Liu, A. C. Potter, K. T. Law, and P. A. Lee, Phys. Rev. Lett. {\bf109}, 267002 (2012).
\bibitem{Xue}Y. Zhang, K. He, C.-Z. Chang, C.-L. Song, L.-L. Wang, X. Chen, J.-F. Jia, Z. Fang, X. Dai, W.-Y. Shan, S.-Q. Shen, Q. Niu, X.-L. Qi, S.-C. Zhang, X.-C. Ma, and Q.-K. Xue, Nat. Phys. {\bf6}, 584 (2010).
\bibitem{CZChen} C.-Z. Chen, J. J. He, D.-H. Xu, and K. T. Law, Phys. Rev. B {\bf96}, 041118 (2017).
\bibitem{Schnyder} A. P. Schnyder, S. Ryu, A. Furusaki, and A. W. W. Ludwig, Phys. Rev. B {\bf78}, 195125 (2008).
\bibitem{Tewari} S. Tewari and J. D. Sau, Phys. Rev. Lett. {\bf109}, 150408 (2012).
\bibitem{PALee} P. A. Lee and D. S. Fisher, Phys. Rev. Lett. {\bf47}, 882 (1981).
\bibitem{Amorim} C. S. Amorim, K. Ebihara, A. Yamakage, Y. Tanaka, and M. Sato,  Phys. Rev. B {\bf 91} 174305 (2015).
\bibitem{Alicea}J. Alicea, Y. Oreg, G. Refael, F. von Oppen, and M. P. A. Fisher, Nat. Phys. {\bf7}, 412 (2011).
\bibitem{Ivanov} D. A. Ivanov, Phys. Rev. Lett. {\bf86}, 268 (2001).
\bibitem{SM} Supplemental Material
\bibitem{WJi}  W. Ji and X.-G. Wen, arXiv:1708.06214 (2017).
\bibitem{YHuang} Y. Huang, F. Setiawan and J. D. Sau, arXiv:1708.06752 (2017).
\bibitem{BLian} B. Lian, J. Wang, X.-Q. Sun, A. Vaezi, and S.-C. Zhang, arXiv:1709.05558 (2017).
\end{thebibliography}
\end{document}